\documentclass{elsarticle}
\usepackage{hyperref}
\usepackage{amssymb}
\usepackage{pifont}
\usepackage{tabularx}
\usepackage{array}
\newcolumntype{C}[1]{>{\centering\arraybackslash}p{#1}}

\journal{Journal of Applied Acoustics}

\usepackage{numcompress}

\begin{document}

\begin{frontmatter}

\title{The Implementation of Low-cost Urban Acoustic Monitoring Devices}


%
\author{Charlie Mydlarz$^{1,2}$, Justin Salamon$^{1,2}$ and Juan Pablo Bello$^{2}$}
\address{$^{1}$Center for Urban Science and Progress, New York University, USA}
\address{$^{2}$Music and Audio Research Laboratory, New York University, USA}


%

\begin{abstract}
The urban sound environment of New York City (NYC) can be, amongst other things: loud, intrusive, exciting and dynamic. As indicated by the large majority of noise complaints registered with the NYC 311 information/complaints line, the urban sound environment has a profound effect on the quality of life of the city's inhabitants. To monitor and ultimately understand these sonic environments, a process of long-term acoustic measurement and analysis is required. The traditional method of environmental acoustic monitoring utilizes short term measurement periods using expensive equipment, setup and operated by experienced and costly personnel. In this paper a different approach is proposed to this application which implements a smart, low-cost, static, acoustic sensing device based around consumer hardware. These devices can be deployed in numerous and varied urban locations for long periods of time, allowing for the collection of longitudinal urban acoustic data. The varied environmental conditions of urban settings make for a challenge in gathering calibrated sound pressure level data for prospective stakeholders. This paper details the sensors' design, development and potential future applications, with a focus on the calibration of the devices' Microelectromechanical systems (MEMS) microphone in order to generate reliable decibel levels at the type/class 2 level.
\end{abstract}


\end{frontmatter}


\section{Introduction \label{sec:intro}}

Noise pollution is an increasing threat to the well-being and public health of city inhabitants \cite{isling2004}. It has been estimated that around 90\% of New York City (NYC) residents are exposed to noise levels exceeding the Environmental Protection Agencies (EPA) guidelines on levels considered harmful to people \cite{hammer2014environmental}. The complexity of sound propagation in urban settings and the lack of an accurate representation of the distribution of the sources of this noise have led to an insufficient understanding of the urban sound environment. While a number of past studies have focused on specific contexts and effects of urban noise \cite{payne2009research}, no comprehensive city-wide study has been undertaken that can provide a validated model for studying urban sound in order to develop long-lasting interventions at the operational or policy level.

To monitor and ultimately foster a greater understanding of urban sound, an initial network of low-cost acoustic sensing devices \cite{mydlarz_RSD_2014} were designed and implemented to capture long-term audio and objective acoustic measurements from strategic urban locations using wireless communication strategies. These prototype sensing devices currently incorporate a quad-core Android based mini PC with Wi-Fi capabilities, and a Microelectromechanical systems (MEMS) microphone. The initial goal is to develop a comprehensive cyber-physical system that provides the capability of capturing, analyzing and wirelessly streaming environmental audio data, along with its associated acoustic features and metadata. This will provide a low-cost and scalable solution to large scale calibrated acoustic monitoring, and a richer representation of acoustic environments that can empower a deeper, more nuanced understanding of urban sound based on the identification of sources and their characteristics across space and time. As part of this goal, work is ongoing to equip the sensors with state-of-the-art machine listening capabilities, briefly discussed in Section~\ref{sec:ssid}, such as automatic sound source identification through the development of novel algorithms. This approach aims to enable the continuous monitoring and ultimately the understanding of these urban sound environments.

\subsection{New York City noise}
In 2014 the NYC 311 information/complaints line \footnote{\url{http://www1.nyc.gov/311/index.page}}, received 145,252 complaints about noise, up 34\% from 2013. As of August 2015, 105,063 noise complaints have already been registered \cite{nyt-noise2015}. NYC has tried to regulate sources of noise since the 1930s and in 1972 it became the first city in the U.S. to enact a noise code \cite{bronzaft2010noise,wu2008toward}. As a result of significant public pressure, a revised noise code went into effect in 2007 \cite{NYC2005}. This award-winning code, containing 84 enforceable noise violations, is widely-considered to be an example for other cities to follow \cite{noisecodeguide}. However, NYC lacks the resources to effectively and systematically monitor noise pollution, enforce its mitigation and validate the effectiveness of such action. Generally, the Noise Code is complaint driven. The NYC Department of Environmental Protection (DEP) inspectors are dispatched to the location of the complaint to determine the ambient sound level and the amount of sound above the ambient, where a notice of violation is issued whenever needed. Unfortunately, the combination of limited human resources, the transient nature of sound, and the relative low priority of noise complaints causes slow or in-existent responses that result in frustration and disengagement.

New York City noise has been the focus of a plethora of studies investigating: noise levels in relation to air pollutants and traffic \cite{ross2011noise,Kheirbek2014}, noise exposure from urban transit systems \cite{neitzel2009noise,neitzel2011exposures,gershon2006pilot} and noise exposure at street level \cite{mcalexander2015street}. All of these highlight the fact that noise is an underrepresented field in urban health and found that average levels of outdoor noise at many locations around the city exceed federal and international guidelines set to protect public health. Sensing of noise conditions using 56 relatively low cost logging sound level meters (SLMs) was investigated in \cite{Kheirbek2014}, where general purpose SLMs were used to log SPL measurements over the period of one week. These type of deployments can help to identify noise patterns over short periods of time with respect to other factors such as traffic intensity, but are lacking in their ability to monitor noise over longer duration's. Long term noise monitoring is required to allow health researchers to perform better epidemiological studies of environmental contributions to cardiovascular disease \cite{weinmann2012subjective}.

With its population, its agency infrastructure, and its ever-changing urban soundscape, NYC provides an ideal venue for a comprehensive study and understanding of urban sound.

\subsection{Type certification and IEC 61672 \label{sec:acoustic-meas}}
In order for a piece of equipment to be suitable for acoustic measurement purposes, it should comply with the sound level meter (SLM) standard IEC 61672-1 \cite{iec61672-1}. This includes, for example, tolerance limits for a device's frequency response, self-generated noise and linearity. Two ``type'' specifications are defined where type 1 devices, designated Precision, are intended for accurate sound measurements in the field and laboratory, type 2 devices, designated General Purpose, are intended for general field use. The overall accuracy of the device is determined by its ``type'' rating. In the US, the general minimum type specification for use in noise surveying is type 2. The American National Standards Institute's 1983 ANSI S 1.4 \cite{ansi-s1.4}  for ``type'' certification shares many similarities with the more recent 2013 IEC 61672-1, however, the later standard does make more demands on the sound level meter regarding accuracy, performance and calibration. It is not the intention of this paper to prove that this sensor network can be used to generate legally enforceable acoustic data for a location, but the data that it can provide will be a real-time, continuous and accurate indication of the acoustic conditions in which each sensor inhabits. This data stream will help to inform and augment urban noise enforcement procedures, e.g. optimizing the allocation of in-depth noise assessment personnel and equipment.

With the current 2013 IEC 61672-1 standard for type ratings, a traditional MEMS microphone does not allow for the full set of test procedures to be carried out. The MEMS diaphragm is electrically connected to the pre-amplifier stage within the microphone's casing which does not allow for the direct injection of an electrical test signal to this internal pre-amplifier as defined in Section 5.1.16 in IEC 61672-1:

\begin{quote}
\textit{5.1.16 The microphone shall be removable to allow insertion of electrical test signals to the input of the pre-amplifier.}
\end{quote}

Thus, MEMS microphones cannot currently be granted a type rating using the 2013 IEC specifications. Future revisions to the standard would surely benefit from an expansion to handle the ever advancing MEMS microphones as the sensing component for low-cost and scalable noise monitoring solutions.

\section{A high quality \& scalable acoustic sensor network \label{sec:sensor_networks}}
The last decade has produced a number of different approaches for environmental noise monitoring. These static acoustic sensor networks vary from expensive, dedicated acoustic monitoring stations to low-cost examples that make use of consumer grade hardware. Advances in low-power computing, microphone technology and networking have provided these dedicated stations incremental upgrades in the form of real-time data transmission capabilities, but these advancements have had the most profound effect on the more flexible low-cost sensor nodes which can now perform advanced DSP (digital signal processing) on audio data captured using high quality microphones and transmit via a number of wireless networking strategies. These statically deployed acoustic sensors can be grouped into three general categories, where sensor functionality and cost are the focus:

\subsection{Category 1 - Dedicated monitoring stations} These commercial devices are designed and built for the purpose of accurate, reliable, low-noise and enforceable acoustic monitoring and as such can cost upwards of \$10,000USD. These systems generally consist of high-end, dedicated portable logging sound level meters and bespoke network hardware, encased in a weatherized housing. An example from this category is the Bruel \& Kjaer 3639-A/B/C \cite{B+K3639}, which retails for $\approx$\$15,000USD and includes a type 1 approved microphone and analysis system with a measurement range from 25-140dBA, the ability to produce 1/3 octave spectral noise data, real-time wireless data transfer, autonomous operation and a ruggedized casing for long term exterior operation. Other examples with similar specifications and price points include the 01dB OPER@ Station \cite{01dboper} and the Larson Davis 831 Noise Monitoring System \cite{larsondavis831nms}. The hardware and software used in these systems is usually proprietary and therefore does not provide the ability to customize the functionality to purposes other than basic acoustic monitoring of noise levels, except through software module purchases such as threshold based event detection typically costing upwards of \$1000 per module. Whilst initial sensor costs are high, maintenance costs are generally less than in lower cost solutions due to the specialized and highly engineered nature of these devices. Deployment durations are generally in the order of a few months at a time due to the high cost of the hardware and security concerns.

\subsection{Category 2 - Moderately scalable sensor network} This group consists of a combination of commercially made and research group developed devices that provide greater opportunities for larger scale deployments than those of Category 1 with varied accuracy of data. The typical price point of each node in this group is the \$600USD mark. Commercial examples include the \$560USD Libelium Waspmote Plug \& Sense, Smart Cities device \cite{libeliumsmartcities} which, amongst other things, measures simple dBA values with no type certification, to give an indication of a location's sound pressure level. The Libelium device is ruggedized for exterior use, runs autonomously, and can transfer data wirelessly to a central server. This system provides no means to process the incoming audio data as the conversion to dBA values occurs at the hardware level on the microphones board. A different example in this category is the RUMEUR network \cite{mietlicki2015} developed by the Noise Observatory Group of the non-profit organization, Bruitparif, based in Paris. Their network consists of around 50 $\approx$\$2500USD monitoring stations gathering high quality audio and accurate acoustic data at the type 1 level, including acoustic event detection. This network is also complemented by 350 $\approx$\$550USD lower-cost devices that log dBA values at the type 2 level. Whilst more scalable than Category 1 networks these are still limited by relatively high costs and in some cases measurement inaccuracies.

\subsection{Category 3 - Low-cost sensor network} This category of sensor network typically consists of custom made nodes designed to be inexpensive, low-power and autonomous for large scale deployments. The majority of these utilize low-power single board computing cores with low-cost audio hardware. The price point of $\approx$\$150 per sensor node in this category make it a viable solution for pervasive network deployments. These networks are currently, predominately developed by university research groups including the MESSAGE project at Newcastle University \cite{bell2013novel}, whose low-cost sensors monitor noise levels in dBA, with an effective range from 55-100dBA at $\approx$3dBA accuracy when compared to a type 1 sound level meter. A similar low-cost initiative from Finland \cite{kivela2011design} has produced sensor nodes costing $\approx$\$150USD that are capable of transmitting dBA values wirelessly using a low-power computing core and audio system capable of an effective range of measurements from 36-90dBA. This category is clearly the more scalable due to its low cost sensor nodes, however, in the examples given, the accuracy of acoustic data is low and the low power computing cores do not allow for any in-situ DSP. 

\subsection{What makes a high quality \& truly scalable acoustic sensor network?}
In order to realize a truly scalable, accurate, autonomous and adaptable system, a combination of attributes from each of the previously mentioned categories is required. Based on these previous examples of acoustic sensing networks, a viable solution for high quality, large scale urban noise monitoring should provide a minimum of these features:

\begin{itemize}
	\item The ability to monitor sound pressure levels with a comparable level of accuracy to city agency standards
	\item Enhanced computing capabilities for intelligent, in-situ signal processing and wireless raw audio data transmission
	\item Autonomous in its operation
	\item A low cost per sensor node at the $\approx$\$100USD price point
\end{itemize}

The presented solution aims to fulfill all of these requirements to provide a viable solution for advanced, large scale urban acoustic monitoring. The proposed sensor nodes will be shown empirically to produce acoustic data at the type 2 level, the high processing power of the computing core will be detailed including its ability to operate autonomously using a combination of components that cost less than \$100USD in parts.

\section{Applications}
Acoustic data gathered using the systems deployed sensor network can be used to identify important patterns of noise pollution across urban settings. Decision makers at city agencies can then strategically utilize the human resources at their disposal, i.e. by effectively deploying costly noise inspectors to offending locations automatically identified by the the proposed network. The continuous and long term monitoring of noise patterns by the network allows for the validation of the effect of this mitigating action in both time and space, information that can be used to understand and maximize the impact of future action. By systematically monitoring interventions, one can understand how often penalties need to be imparted before the effect becomes long-term. With sufficient deployment time, 311 noise complaint patterns could also be compared to the network's data stream in a bid to model and ultimately predict the occurrence of noise complaints. The overarching goal would be to understand how to minimize the cost of interventions while maximizing noise mitigation. This is a classic resource allocation problem that motivates much research on smart-cities initiatives, including this one.

The eventual increase in network deployment across large urban areas will allow for noise mapping with high spatial and temporal resolution. Examples of the long term goals accomplishable with this approach and the use of existing geo-located datasets include: how sound impacts on the health of a city's population, correlates with urban problems ranging from crime to compromised educational conditions, and how it affects real estate values.

\section{Summary of contributions}

This paper details the design and measurement of a low-cost MEMS solution for a novel acoustic sensing device. These specific contributions are made to the field of noise monitoring in smart cities:

\begin{itemize}

\item Measurements as per the IEC 61672-1 specification for sound level meters show the suitability of an analog MEMS microphone solution for accurate urban acoustic monitoring at the type 2 level

\item The use of consumer mini PC devices in acoustic sensing devices allow for advanced signal processing to be performed in-situ for applications such as automatic sound source classification

\item The low cost of the core components of the proposed sensor device provide an advanced and scalable system for acoustic sensing in smart cities

\end{itemize}

The paper begins by focusing on the core hardware components of the sensor device, followed by the measurement process carried out on the proposed MEMS microphone solution. It concludes with a summary of the findings and a discussion of the future work. As the main focus is on the hardware development and testing, the sensor networks software and networking elements have been omitted in this paper.

\section{Hardware \label{sec:hw}}
\subsection{Computing core}
\begin{figure}[h]
\begin{center}
\includegraphics[width=0.4\textwidth]{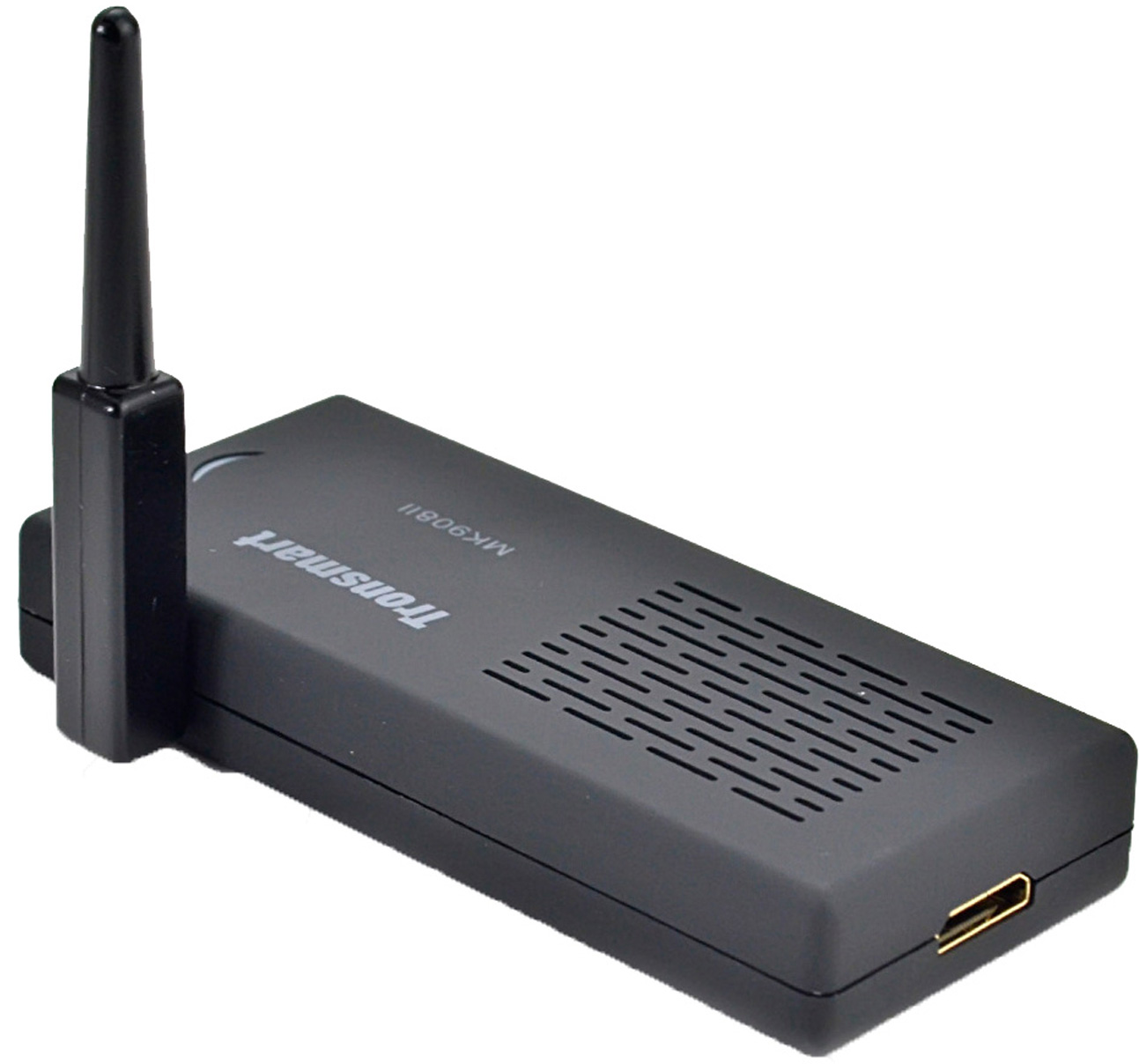}
\end{center}
\caption{Tronsmart MK908ii mini PC\label{fig:tronsmart}}
\end{figure}

The proposed sensing node is based around a consumer computing platform where low cost and high power are of paramount concern. The design philosophy is based on the creation of a network that provides dense spatial coverage over a large area, through the deployment of inexpensive and physically resilient sensors, whose housing considerations are included in \cite{mydlarz_RSD_2014}. At the core of the sensing device is a single board Tronsmart MK908ii mini PC running a Linux Ubuntu 13.04 based operating system. These small and versatile devices shown in Figure~\ref{fig:tronsmart} are priced at \$50USD as of August 2015 and provide a 1.6GHz quad core processor, 2GB of RAM, 8GB flash storage, USB I/O, and Wi-Fi connectivity. The computing power offered by these units allows for complex digital signal processing to be carried out on the device, alleviating the need to transmit large amounts of audio data for centralized processing.

These mini PCs provide an all in one computing solution that incorporates a number of ready to use Wi-Fi and flash storage components in a small package. In contrast, other mini PC solutions, such as the ODROID C1+ from Hard Kernel \cite{odroid2015}, the BeagleBone Black \cite{beagleblack2015} and the Raspberry Pi 2 Model B \cite{rpi2015}, retail at \$35-\$55, but at that price do not include a suitable USB Wi-Fi module or any flash storage. These items must be purchased separately. However, when purchased in bulk these other devices may become viable solutions in terms of cost. Table \ref{tab:computing-compare} compares the mini PC used to similar solutions as purchased.

\begin{table}[h]
\centering
\small
\renewcommand{\arraystretch}{0.75}
\begin{tabularx}{\textwidth}{l C{0.7cm} c C{2.0cm} C{1.0cm} C{1.0cm}}
\hline
\textbf{Mini PC}	& \textbf{Cost (USD)}	& \textbf{Cortex CPU}	& \textbf{RAM}	& \textbf{Storage}	& \textbf{Wi-Fi}\\
\hline
ODROID C1+				& 37		& 	A5 1.5GHz 4 core	& 1GB DDR3 & \ding{55} 		& \ding{55}	\\
Raspberry Pi 2B			& 35		& 	A7 0.9GHz 4 core	& 1GB DDR2 & \ding{55} 		& \ding{55}	\\
BeagleBone Black			& 55		& 	A8 1.0GHz 1 core	& 0.5GB DDR3 & 4GB 	& \ding{55}	\\
Tronsmart MK908ii		& 50		& 	A9 1.4GHz 4 core	& 2GB DDR3 & 8GB 	& \ding{51}\\
\hline
\end{tabularx}
\caption{Comparison of ODROID C1+, Raspbery Pi 2 Model B, BeagleBone Black and Tronsmart MK908ii\label{tab:computing-compare}}
\end{table}

The similar mini PCs currently available contain a less powerful CPU and reduced RAM making them less amenable for advanced digital signal processing (DSP) applications such as automatic sound source classification. Based on this comparison, the Tronsmart MK908ii provides a more complete solution for high quality acoustic sensing applications owing to its superior processing power, RAM, inbuilt storage and Wi-Fi module. However, with the constant development and subsequent increase in computational power of these single board computers, solutions such as the Raspberry Pi may become viable solutions in terms of processing capability in the near future.

USB I/O allows for the inclusion of a USB audio device to handle all analog to digital conversion (ADC) work, thus providing the means to connect a custom microphone solution. The USB audio device chosen for this application had to be: compatible with Linux based devices, low in price, provide input gain control and a clean signal path. The device selected was the eForCity USB audio interface which retails for \$5USD as of August 2015. It provides a single microphone input channel with low noise and a software adjustable input gain stage.

\begin{figure}[h]
\begin{center}
\includegraphics[width=1.0\textwidth]{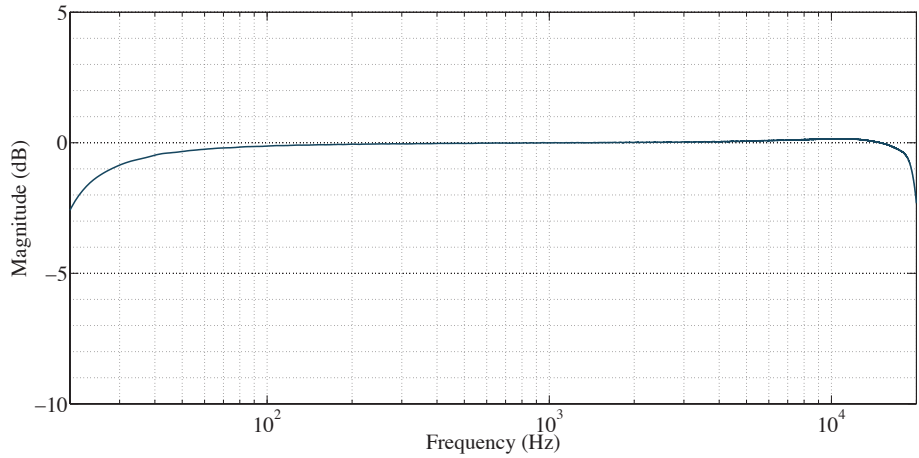}
\end{center}
\caption{eForCity USB audio interface frequency response (20Hz-20kHz) normalized at 1kHz\label{fig:eforcity-resp}}
\end{figure}

The frequency response of the device was measured and whilst it introduces filtering with a steep roll-off below 20Hz and above 20kHz, the audible frequency range is relatively unaffected. Figure \ref{fig:eforcity-resp} shows this response graphically. The measured noise floor of the device with 0dB of gain applied was \textbf{-90.1dBV(A)}, providing a wide dynamic range for urban acoustic measurement.

\subsection{MEMS microphones \label{sec:mems-mics}}
In recent years, interest in MEMS microphones has expanded due to their versatile design, greater immunity to radio frequency interference (RFI) and electromagnetic interference (EMI), low cost and environmental resiliency \cite{van2011ability,barham2009development,barham2010practical}. This resiliency to varying environmental conditions is particularly important for long term acoustic monitoring applications in the harsh subzero winters and hot and humid summers of NYC. A study characterizing a custom MEMS microphone solution for acoustic measurement purposes \cite{scheeper2003new} exhibited a very low temperature coefficient for sensitivity of $<$0.017dB/$^{\circ}$C. A large variation in humidity was also shown to have a minimal impact on the MEMS microphones sensitivity, with decreases of $<$0.1dB between relative humidity (\%RH) conditions of 40\% and 90\%. 

Current MEMS models are generally 10x smaller than their electret counterparts. This miniaturization has allowed for additional circuitry to be included within the MEMS housing, such as a pre-amp stage and an ADC to output digitized audio in some models. The production process used to manufacture these devices also provides an extremely high level of part-to-part consistency in terms of acoustic characteristics such as sensitivity and frequency response, making them more amenable to multi-capsule and multi-sensor arrays, where consistency of individual microphones is paramount.

\begin{figure}
\begin{center}
\includegraphics[width=0.6\textwidth]{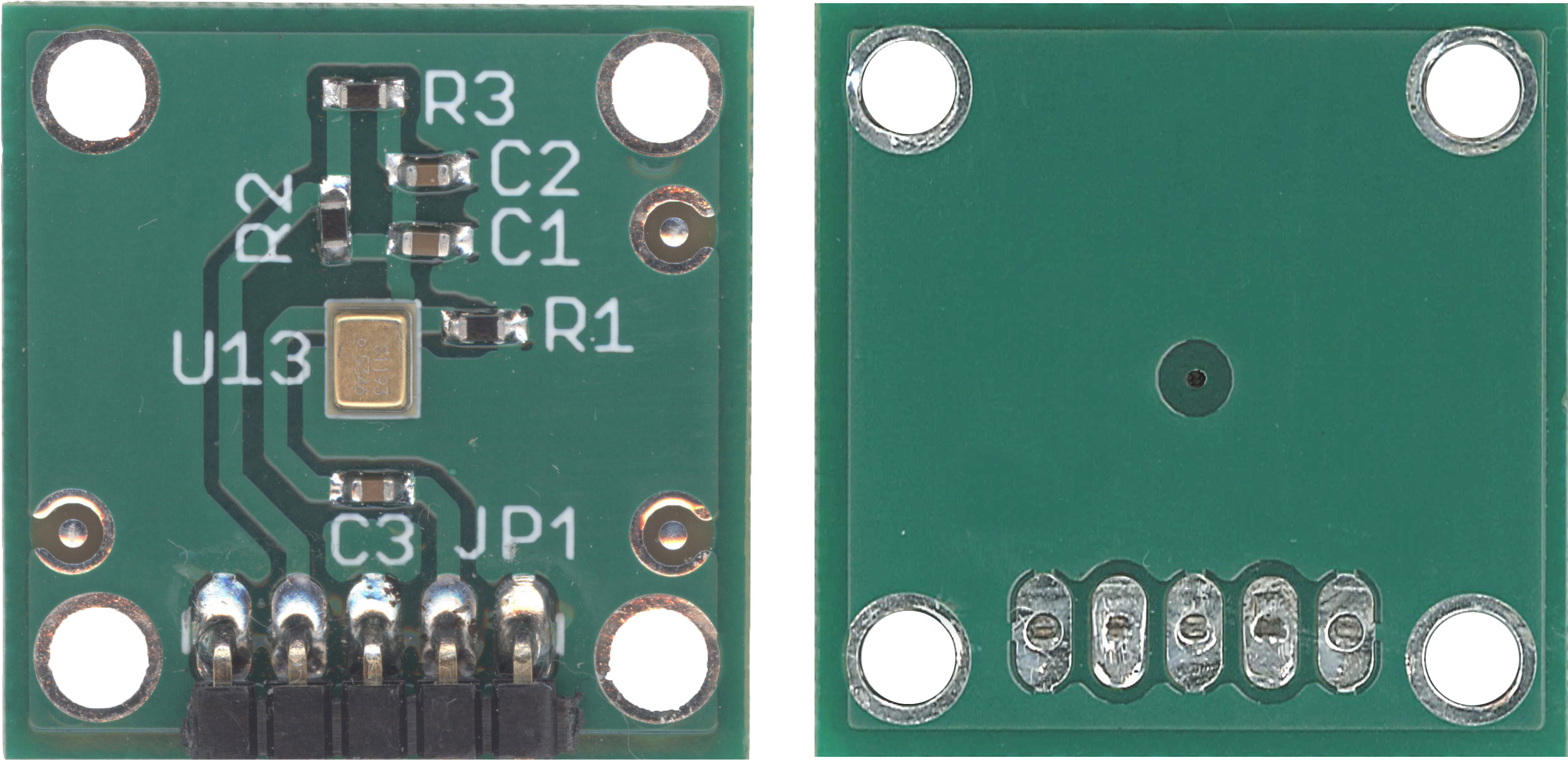}
\end{center}
\caption{Front/back MEMS microphone custom PCB (Knowles SPU0410LR5H-QB microphone in center of left image)\label{fig:MEMS_board}}
\end{figure}

In the proposed prototype microphone system we investigate the Knowles SPU0410LR5H-QB. The silicone diaphragm MEMS microphone has a manufacturer quoted ``flat frequency response'' between 100Hz and 10kHz. It requires a maximum 3.6V supply and draws only 120$\mu\mbox{A}$. In addition, it is quoted as having a sensitivity of -38dB re.~1V/Pa and a signal-to-noise ratio of 63dBA. In order to test the Knowles MEMS microphone a PCB shown in Figure~\ref{fig:MEMS_board} was designed and fabricated \cite{mydlarz_MEMS_2014}. It was found in testing that the switched mode power supply noise created by the low-cost AC-DC converters used to power the MEMS was unnecessarily high, see Section~\ref{sec:psu_consid}. To reduce this to acceptable levels an LT1086 linear voltage regulator was introduced to reduce the noisy USB 5V down to a clean 3.6V DC supply. The use of adequately shielded audio cabling is also crucial in this application as the low-level audio signal from the MEMS microphone board is running in close proximity to the radio frequency (RF) components of the mini PC. This RF interference (RFI) has been observed on an un-shielded version of the system as a low-level broadband noise burst at times of high Wi-Fi activity. A proper shielding and grounding strategy reduces this RFI noise but does not remove it entirely from the signal chain. The test results in this paper were gathered using the audio components in isolation with no RF components present using the configuration described above. The total cost of the components used in the solution is around \$7USD as of August 2015.

\subsection{Microphone mount}

\begin{figure}
\begin{center}
\includegraphics[width=0.6\textwidth]{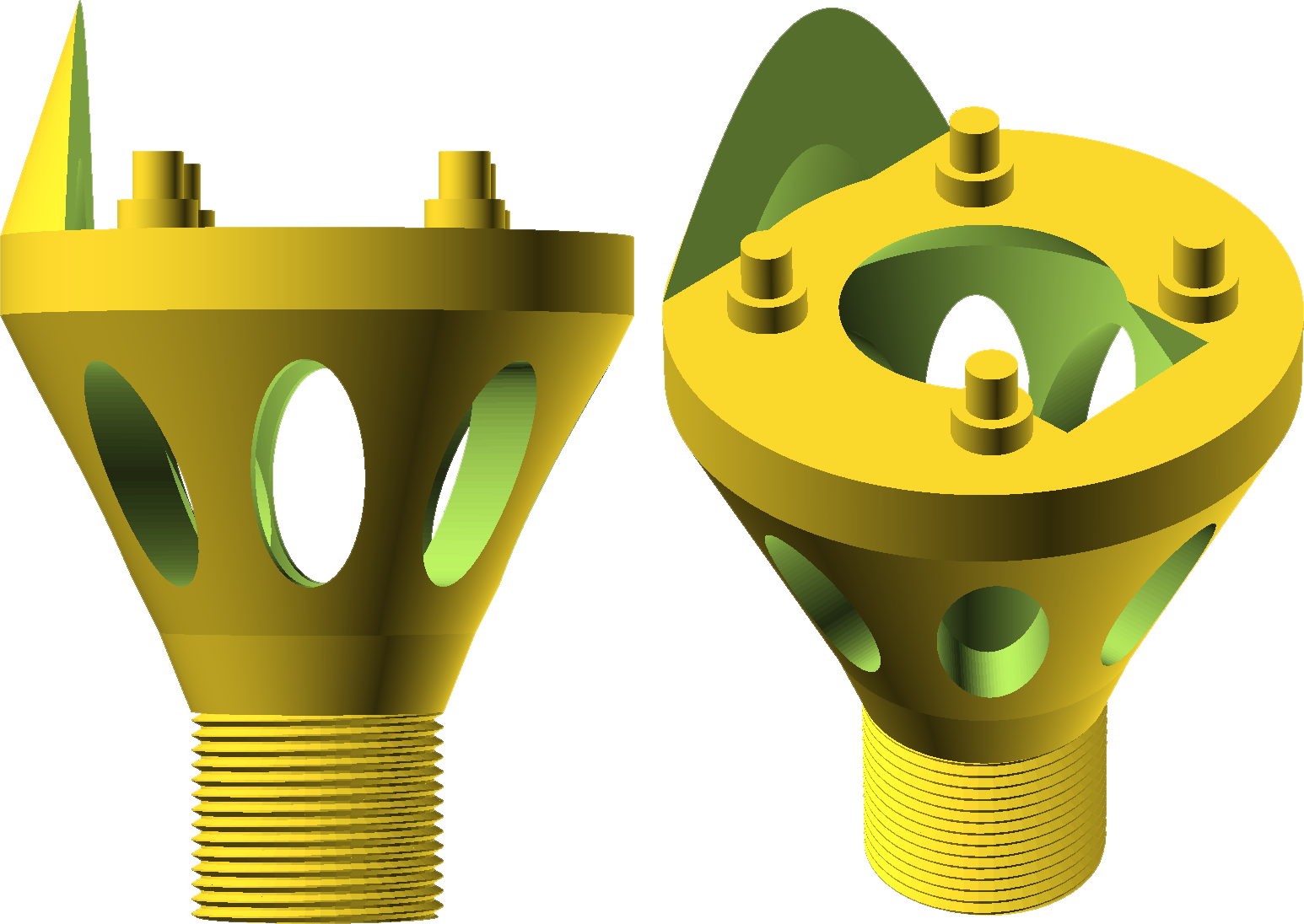}
\end{center}
\caption{Custom 3D printable microphone mount with four microphone mount pegs shown on top\label{fig:mic-mount}}
\end{figure}

In order to securely mount the MEMS microphone board a custom ABS plastic mount was fabricated. This 3D printed component is shown in Figure \ref{fig:mic-mount} and ensures the microphone port is unobstructed, protected from water droplets due to the protruding lip and allows for a windshield to be placed around the mount to reduce the effects of wind noise on the microphone.

The open space behind the microphone board mount point ensures no Helmholtz resonances can build up as a result of a closed cavity close to the microphone. Complex diffraction effects from off-axis sound sources may have an effect on the response at frequencies of  $>$8.5kHz, which corresponds to the 40mm diameter of the custom microphone mount. The dimensions and shape of the MEMS microphone PCB also have the potential to effect response at the  $>$13.5kHz range. These effects will be investigated in a further stage of testing as mentioned in Section~\ref{sec:conclusions}. The top pegs allow the microphone board to be securely seated, reducing the chance of any mechanical rattling. Externalizing the microphone board in this way also reduces the effects of RFI from the mini PC's Wi-Fi module located within the sensors aluminum housing.

\subsection{Power supply considerations\label{sec:psu_consid}}

The current sensor design utilizes a constant connection to a 120V mains supply via a domestic power outlet. One of the main sources of unwanted noise in the audio signal chains stems from the audio systems power supply unit or PSU. The key to recording ``clean'' analog signals is to provide ``clean'' power to the audio system. Any AC noise present on the DC supply of an audio component will be transferred, to some degree, into the analog audio signal. In the presented low-cost acoustic sensor a single PSU supplies the 5V DC supply for the mini PC, which in turn supplies the analog MEMS microphone its 3.6V DC. A significant source of noise in a sensor such as this is load transients, which are caused by sudden, large current drains from the mini PC's Wi-Fi module and CPU. These produce ringing on the power rails which make their way into the audio signal if not properly dealt with. A \$3USD switched mode PSU was measured using an oscilloscope after removing its 5V DC component and can be seen alongside its voltage regulated signal using the LT1086 linear voltage regulator.

\begin{figure}
\begin{center}
\includegraphics[width=1.0\textwidth]{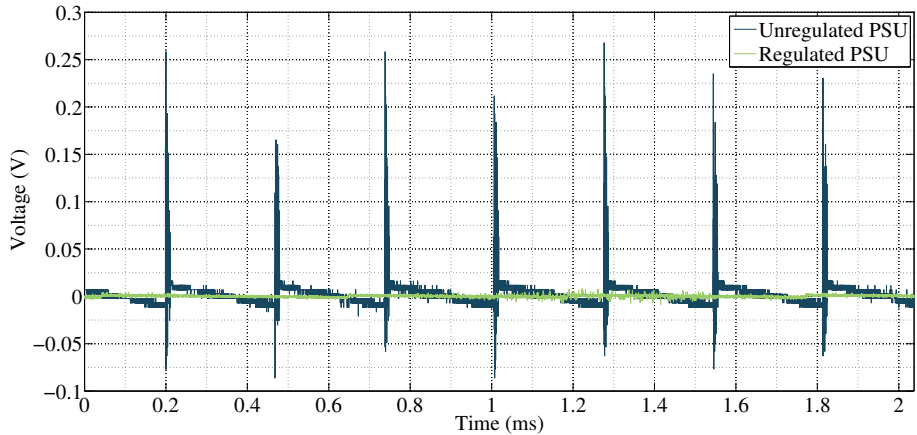}
\end{center}
\caption{Comparison of DC supply noise of unregulated and regulated \$3USD PSU (5V DC offset removed)\label{fig:psu_compare}}
\end{figure}

Figure \ref{fig:psu_compare} shows the high levels of noise present on the unregulated PSU. Average peak-peak levels of 350mV were observed. These pulses are the result of the switching frequency of the switched mode power supply. The regulated PSU signal shows a vastly improved noise level with a peak-peak average of 17mV.

\begin{figure}
\begin{center}
\includegraphics[width=1.0\textwidth]{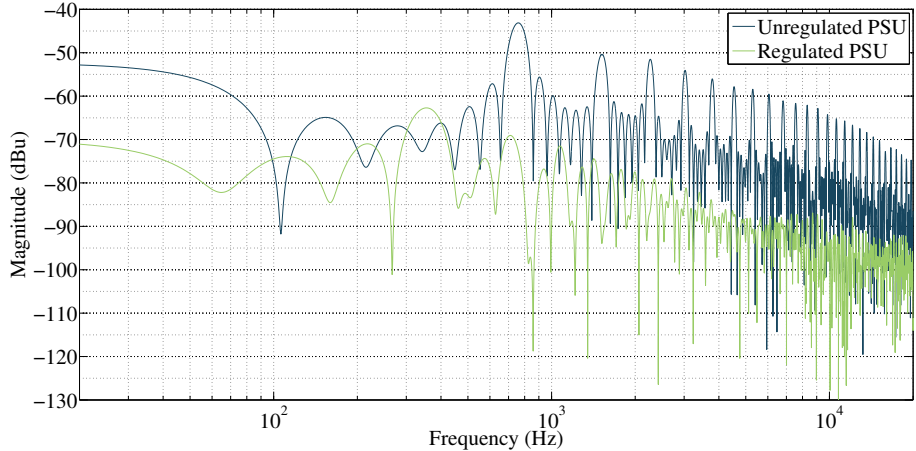}
\end{center}
\caption{Magnitude spectrum (20Hz-20kHz) of unregulated and regulated \$3USD DC power supplies, AC noise (5V DC offset removed)\label{fig:psu_compare_spec}}
\end{figure}

Figure \ref{fig:psu_compare_spec} shows the magnitude spectrum of these DC supply signals. The unregulated supply has a large amount of harmonic noise caused by the switching of the PSU, with its fundamental peak well within the audible range at around 750Hz. The regulated version shows that this high level harmonic content has been greatly attenuated with reductions of upto 26dBu at certain frequencies. Power supply conditioning using grounded capacitors on the DC supply can help in reducing this parasitic AC noise, but in conditions where load transients are also occurring due to Wi-Fi and CPU activity, an additional voltage regulator can provide a low-cost, consistent and ``clean'' DC supply for high quality audio recording.

\subsection{Form factor \& cost of parts}
\begin{figure}
\begin{center}
\includegraphics[width=1.0\textwidth]{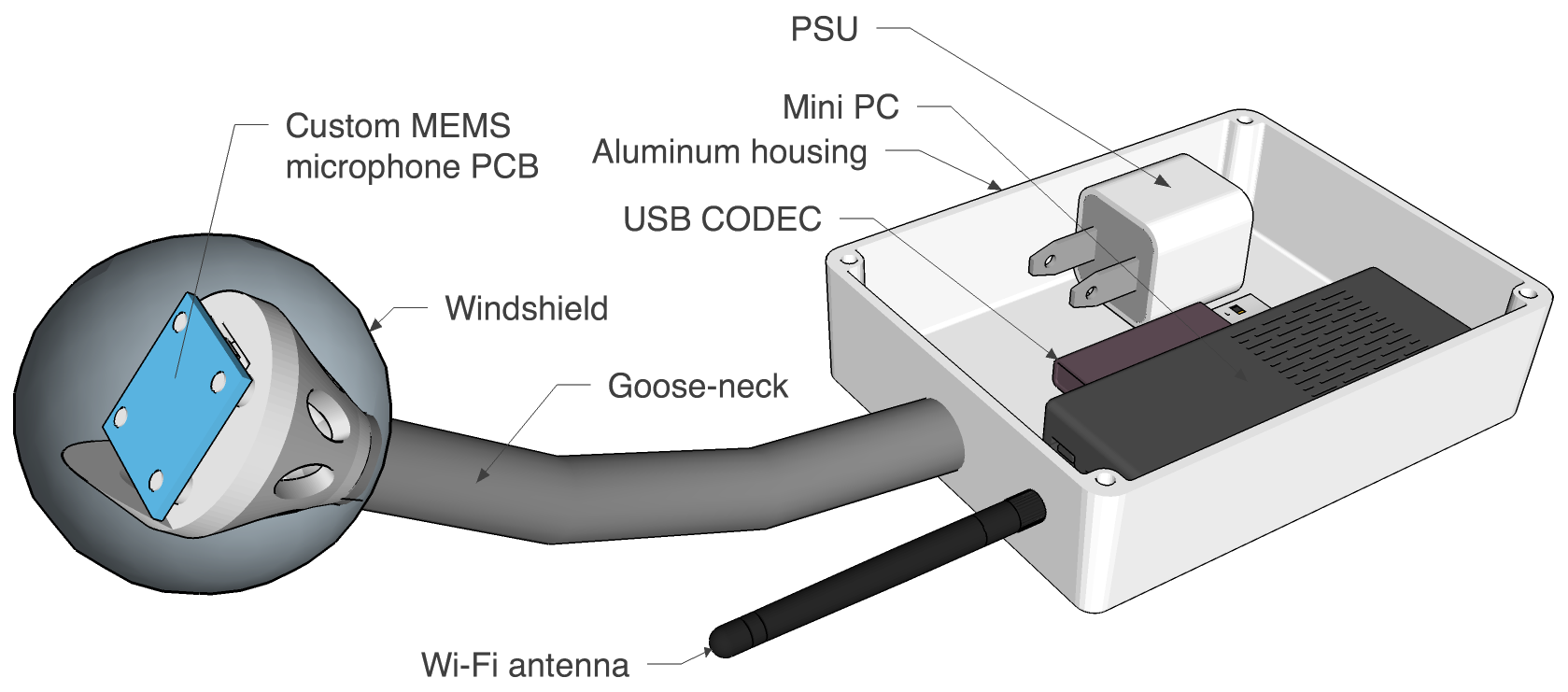}
\end{center}
\caption{Prototype open acoustic sensor node showing core components viewed from the underside\label{fig:xray_sensor}}
\end{figure}

The sensor's prototype housing and form factor is shown in Figure \ref{fig:xray_sensor}. The low-cost unfinished/unpainted aluminum housing was chosen to reduce RFI interference from external sources, solar heat gain from direct sunlight \cite{hoffman2015} and it also allows for ease of machining. All of the sensor's core components are housed within this rugged case except for the microphone and Wi-Fi antenna which is externalized for maximum signal gain.

In the prototype node shown in Figure \ref{fig:xray_sensor}, the MEMS microphone is mounted externally via a flexible but rigid metal goose-neck allowing the sensor node to be reconfigured for deployment in varying locations such as building sides, light poles and building ledges. Acoustic testing of the entire enclosure with the microphone board mounted with its windshield will be carried out when the prototype is in a more advanced stage of production.

\begin{table}[h]
\centering
\small
\renewcommand{\arraystretch}{0.75}
\begin{tabular}{rl}
\hline
\textbf{Component} 			& \textbf{Cost (USD)}\\
\hline
Mini PC 			& 50 		\\
Housing			& 8			\\
Goose-neck		& 6			\\
MEMS mic. board	& 5			\\
USB CODEC		& 5			\\
PSU				& 3			\\
Cabling			& 3			\\
Windshield		& 1			\\
\cline{2-2}
~			& \textbf{81}\\
\cline{2-2}
\end{tabular}
\caption{Core component list \& costs (as of August 2015) for prototype sensor node\label{tab:parts-costs}}
\end{table}

The total cost of core parts for the prototype sensor node is broken down in Table \ref{tab:parts-costs}, with the items ordered by descending price. The total cost of parts excludes construction and deployment costs, but is very low for such a capable system when compared to similar acoustic sensing nodes.

\section{Software \& network}
The sensor nodes software \& network aspects will be briefly summarized for its initial configuration of high quality raw audio capture.
\subsection{Raw audio capture}
The presented sensor node continuously samples 16bit audio data at 44.1kHz. If remote raw audio data collection is required, contiguous one minute segments of audio are first compressed using the lossless FLAC audio encoder \cite{flac2015}. This FLAC file is encrypted using 128bit Advanced Encryption Standard (AES) encryption, with the AES password encrypted using the RSA public/private key-pair encryption algorithm, resulting in a file that cannot be decrypted unless you are in possession of the private key which only resides on the project's central server. The original raw audio files are removed. The encrypted files can be stored on the device as an additional backup and removed as needed when storage space is running low. The on-board flash storage of the mini PCs allow for up-to 2 days of compressed and encrypted 16bit/44.1kHz audio data to be stored. Any DSP required would be carried out prior to this stage on shorter length audio buffers.

\subsection{Network control}
In the prototype configuration, each sensor node communicates directly with an internet connected Wi-Fi router for data transmission and sensor communication/control. The sensor node uploads audio data at 1 minute intervals. With each of these transmissions the server can respond with a command that the node should carry out. Examples of these commands could be a: data flush request to clear out existing backup audio data, device reboot, manual microphone gain adjust or software update.

\section{Signal pre-processing}
\subsection{Frequency response compensation \label{sec:freq-resp}}
\begin{figure}[h]
\begin{center}
\includegraphics[width=1.0\textwidth]{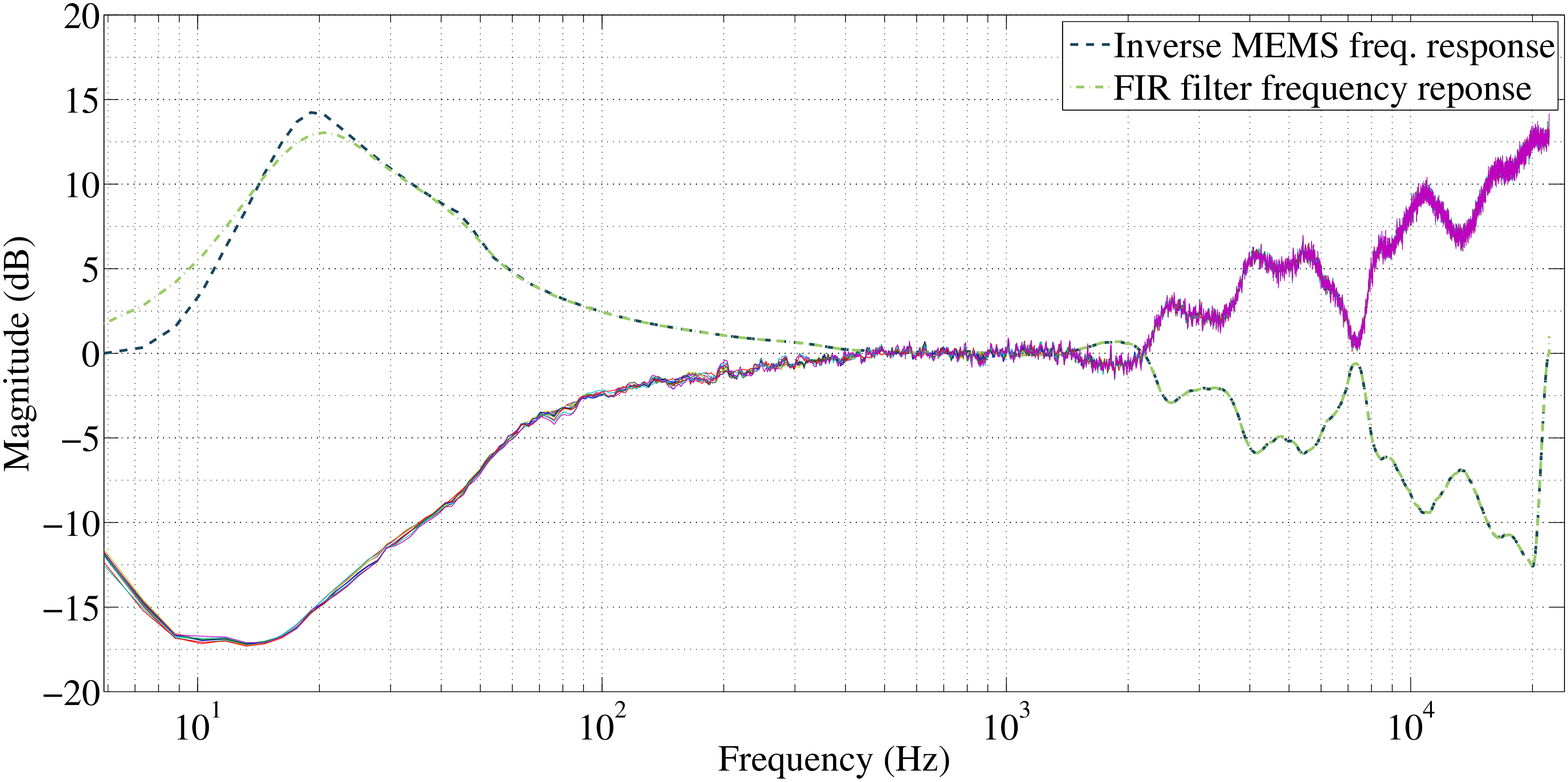}
\end{center}
\caption{MEMS frequency response of 10 microphones (clustered lines) showing consistency between microphone capsules \& regularized compensation filter response with corrospsonding FIR filter response using 8192 coefficients (dashed lines)\label{fig:mems-resp-average}}
\end{figure}

The MATLAB toolbox: Scan IR \cite{boren2011multichannel} was used to generate the impulse responses of the reference microphone and MEMS microphone (referred to as the device under test, DUT) using the swept sine technique. The signals were reproduced through a studio quality Mackie HR824 active speaker and a reference PCB 377B02 microphone and PCB 426E01 pre-amplifier (assumed to be flat in frequency response from 20Hz-20kHz) were used to subtract the room and speaker coloration from the DUT's impulse response. Reference and DUT microphones were placed at 1m from the center point of the speaker on-axis, 1.3m from the floor. The DUT impulse response was generated from an average of 10 microphone boards, whose frequency response are overlaid in Figure \ref{fig:mems-resp-average}. Maximum observed differences between MEMS response's were calculated at 1.0dB, with an average standard deviation between responses of 0.1dB.

As is evident in Figure \ref{fig:mems-resp-average}, negligible differences were observed in frequency response between the 10 MEMS microphones, highlighting the part-to-part consistency of these devices. The peaks and troughs in sensitivity between 2-20kHz could be partly explained by the microphone mounting conditions. The PCB the microphone is mounted to may develop resonances that would reside in this frequency range and result in these observed effects. The rise in response after 10kHz, however, is a result of the Helmholtz resonance created by the microphone's inner chamber and PCB port \cite{Weigold2006}. This averaged response was then used to design an inverse linear-phase FIR filter that would allow for the time-domain filtering of any test signals captured by the DUT, compensating for the MEMS microphone response. The inverse filter was regularized to prevent the filter from applying extreme attenuation or amplification at the high and low frequency ranges as can be observed in the dashed filter response line in Figure \ref{fig:mems-resp-average} at 20Hz and 20kHz. The process was adapted from \cite{bouchard2006inverse}, where a tapered window between 0 and 1 is applied to the high and low extremes of the desired inverse frequency response before the FIR filter is designed. The resultant 8192 coefficient filter provides a close match to the desired response at lower frequencies. This can be efficiently implemented using the optimized DSP routines of the mini PC's Cortex A9 processor \cite{neon2015} providing compensation for the MEMS microphone response in real-time, allowing for the unbiased, in-situ calculation of dBA levels. This regularization process also ensures no sub-sonic frequency content is unnecessarily amplified, improving the systems overall signal to noise ratio. However, the filter gain applied at frequencies between 20-400Hz may serve to increase the overall noise floor of the system, which will be revealed when the self generated noise is quantified in Section~\ref{sec:sgn}.

\subsection{Calibration \label{sec:cal}}
\begin{figure}[h]
\begin{center}
\includegraphics[width=0.5\textwidth]{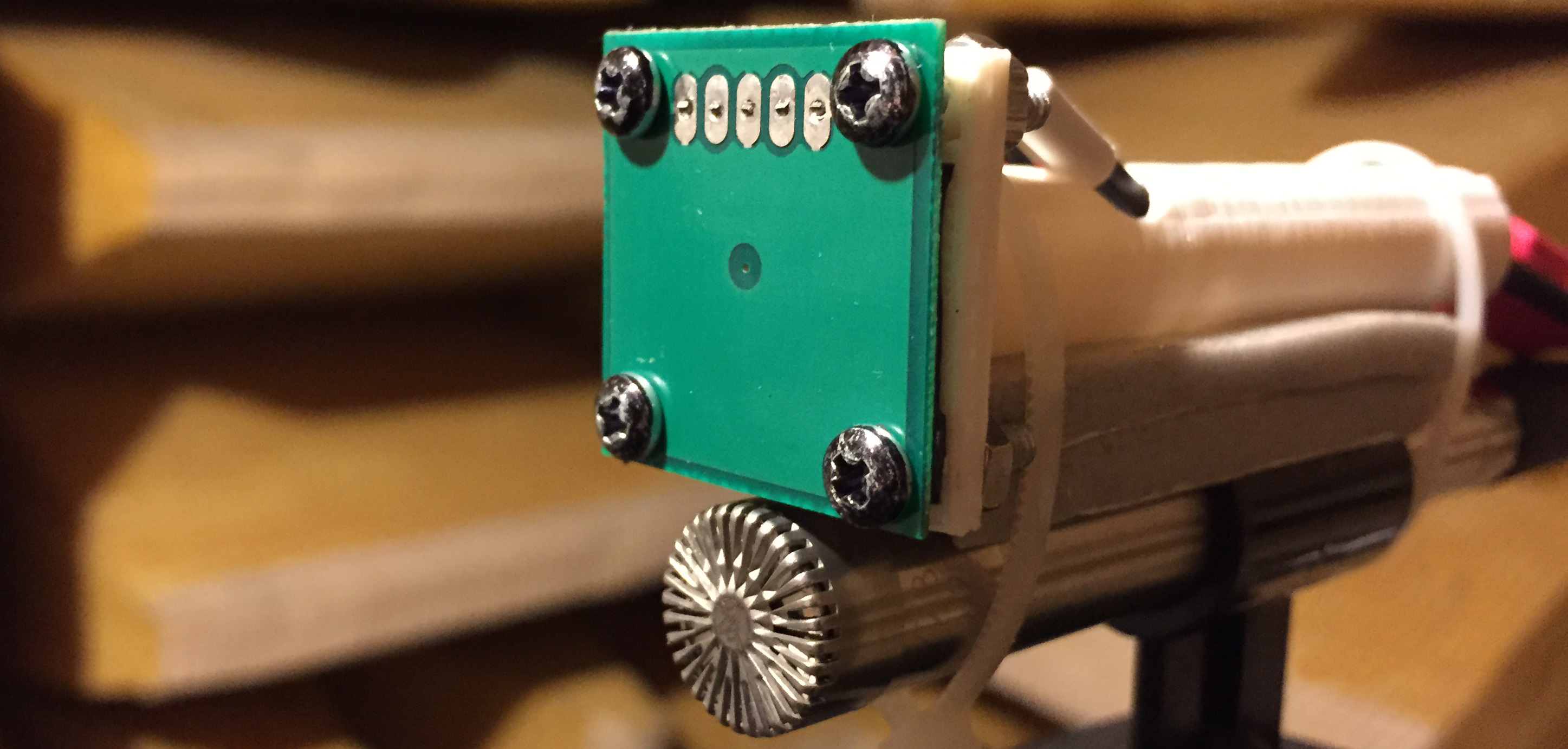}
\end{center}
\caption{\label{fig:mount}DUT (top) and SLM (bottom) microphones mounted}
\end{figure}

The DUT was mounted directly beside the calibrated reference SLM microphone, shown in Figure~\ref{fig:mount}. The devices were positioned at a height of 1.3m and at a distance of 1m on-axis from the center point of the speaker.

\begin{figure}[h]
\begin{center}
\includegraphics[width=40mm]{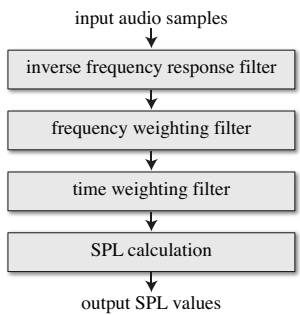}
\end{center}
\caption{\label{fig:block}Block diagram of sensor's SLM functionality}
\end{figure}

The distance between the center of each microphone capsule is 20mm, which was found to produce negligible ($<$0.1dBA) variations in level response when the SLM microphone's position was shifted to match that of the DUT. The output sound pressure level in dBA from the DUT is calculated from the A weighting filtered sample values, which represent the AC voltage produced when presented with the calibration signal of a 1kHz sine wave at 94dBA. An offset adjustment is then applied in order to match the 94dBA SPL input level. Figure~\ref{fig:block} shows the processes required to generate the calibrated SPL output from the DUT.

\section{Measurements \label{sec:meas}}
In order to determine the proposed device's ability to generate type 2 sound pressure level (SPL) data, the device was subjected to a subset of the IEC 61672-3 \cite{iec61672-3} acoustical test procedures, which describe the international standards for periodic testing of SLMs. IEC 61672-1 \cite{iec61672-1} provides the criteria for determining a complete SLM's ability to act as a type 1 or 2 device, including its directivity, which will be affected by the device and microphone housing. This extended set of tests will be performed on the final prototype sensor device in a more advanced stage of its development.

In the following set of measurements the SLM output (Larson Davis 831 - calibrated at the beginning of each measurement stage using the type 1 Larson Davis CAL200) will be used as a reference for comparison to the DUT readings to assess its ability to produce type 2 data. As the SLM is a type 1 certified device, it has its own set of inaccuracies associated with it. It has met the type 1 specifications within the defined tolerance bounds for that standard, thus for the DUT to meet the type 2 specifications, the type 1 tolerance bounds must be factored into the DUT assessment. For example, if the type 2 tolerance bounds for a particular measurement response are $\pm$2.0dB with the corresponding type 1 bounds at $\pm$1.0dB, the adjusted acceptable bounds for the type 2 class in this instance are $\pm$1.0dB (type 2 tolerance range of 4dB minus the type 1 range of 2dB) when using the SLM as the reference device. These will be referred to as the ``adjusted tolerances''. All of the following output values were generated from an average of 4 repeat measurements, where none of the test equipment was moved or altered. No discernible variations ($<$0.1dB) in output were observed between the individual measurements before averaging.

Measurements were conducted under low level ($<$20dBA), fully anechoic conditions at the Cooper Union, Vibration and Acoustics Laboratory~\footnote{\url{http://www.cooper.edu/engineering/facilities/mechanical-engineering/vibration-and-acoustics}}. The atmospheric conditions in the anechoic chamber were measured at the beginning and end of the measurement process ($\approx$2 hrs), and varied from 22-24$^{\circ}$C in air temperature and 50-55 \%RH in relative humidity.

\subsection{Self generated noise \label{sec:sgn}}
The DUT's self generated noise (IEC 61672-1/5.7) was measured under low level, fully anechoic conditions, with all noise generating test equipment located outside of the chamber. Throughout the duration of the 60s measurement period, the reference SLM logged an average SPL of 22.5dBA, close to its lower limit of 19dBA. The self generated noise of the DUT was measured at \textbf{29.9dBA} (max. 30.1dBA, min 29.7dBA, std. 0.1dBA). The dynamic range was then calculated using the manufacturer quoted acoustic overload point of the MEMS microphone. This results in an effective dynamic range of \textbf{88.1dBA}, with an acoustic overload point of \textbf{118dBA}. The signal to noise ratio (94dBA @ 1kHz) of the system was measured at \textbf{64.1dBA} (max. 64.9dBA, min 63.7dBA, std. 0.3dBA). The 29.9dBA noise floor of the system could be partly attributed to the frequency response compensation filter outlined in Section~\ref{sec:freq-resp}. The filter gain at low frequencies brings up the noise floor of the system due to the low frequency roll-off of the analog MEMS microphone. The use of a MEMS microphone with a closer to flat response should serve to mitigate this problem as there will be less reliance on the need to compensate for reduced sensitivity at low frequencies.

The self generated noise value determines the minimum SPL the system can reliably detect. For an urban acoustic sensor in the relatively loud conditions of NYC this level is well below even a quiet suburban setting \cite{united1974information}. The World Health Organization (WHO) night noise guidelines for Europe \cite{worldnight} state that outdoor levels of 30dBA show no observed health effects on humans. The dynamic range value calculated is more than adequate for the acoustic measurement of urban sound environments.

The high end category 1 sensors discussed in Section \ref{sec:sensor_networks} typically exhibit self generated noise levels of around 20dBA with dynamic ranges of around 115dBA common place. Category 3 devices however have been shown to perform far worse than the presented system with dynamic ranges of around 50dBA.

\subsection{Acoustical signal tests of a frequency weighting \label{sec:freq-weighting}}
To test the DUT's ability to produce accurate dBA output for different frequencies (IEC 61672-1, 5.5), it was mounted as in Section~\ref{sec:cal} and subjected to a test signal comprised of 9 steady state 20s sine waves, separated with 5 seconds of silence at octave frequencies from 31.5Hz to 8kHz. Table~\ref{tab:clause_12} shows the mean dBA response from the reference SLM, the DUT, the difference between these two and the adjusted tolerance limits for type 2 devices as discussed at the beginning of Section~\ref{sec:meas}. Standard deviations of the DUT measurements were $<$0.1dBA at all frequencies.

\begin{table}[h]
\centering
\small
\renewcommand{\arraystretch}{0.75}
\begin{tabular}{ccccc}
\hline
Freq. (Hz) & DUT & Ref. & $\Delta$ & Adj. tol.\\
\hline
31.5 & 44.8 & 45.2 & 0.4* & $\pm$ 1.5 \\

63 & 63.6 & 63.7 & 0.1* & $\pm$ 1.0 \\

125 & 76.6 & 76.2 & 0.4* & $\pm$ 0.5 \\

250 & 85.3 & 84.9 & 0.4* & $\pm$ 0.5 \\

500 & 90.2 & 89.9 & 0.3* & $\pm$ 0.5 \\

1k & 93.9 & 94.0 & 0.1* & $\pm$ 0.3 \\

2k & 93.6 & 94.2 & 0.6* & $\pm$ 1.0 \\

4k & 94.1 & 93.3 & 0.8* & $\pm$ 2.0 \\

8k & 93.2 & 90.6 & 2.6* & $\pm$ 3.0 \\

pink & 79.9 & 80.0 & 0.1 & N/A \\

white & 87.5 & 88.0 & 0.5 & N/A \\
\hline

\end{tabular}
\caption{{Acoustical signal tests in mean dBA, varying frequency (* indicates IEC61672-1 criteria met)\label{tab:clause_12}}}
\end{table}

The DUT met all of the adjusted type 2 criteria for dBA frequency weightings when compared to the type 1 SLM. In addition, the response of the DUT and SLM were compared for a 20s, continuous level pink and white noise signal, showing a maximum difference in response of \textbf{0.5dBA}.

\subsection{Long-term stability \label{sec:lts}}
In order to test the long term stability of the DUT, it was subjected to a 30min 1kHz sine wave at 94dBA. The measured difference between the dBA reading at the beginning and end of this period must be within the type 2 tolerance of $\pm$0.2dBA stated in IEC 61672-1, 5.14. The DUT met this criteria, with an observed difference of \textbf{0.07dBA} with mean and standard deviation values throughout the measurement period of $<$0.1dBA.

\subsection{Level linearity \label{sec:level-lin}}
The DUT was subjected to sine waves, linearly increasing up to 94dBA in level to test for the devices linear response to varying SPL's at different frequencies (31.5Hz - 8kHz in octave increments). This was carried out using an acoustical signal under anechoic conditions to test the entire systems response, as opposed to introducing an electrical signal directly into the pre-amp as per IEC 61672-1, 5.6.

\begin{figure}[h]
\begin{center}
\includegraphics[width=1.0\textwidth]{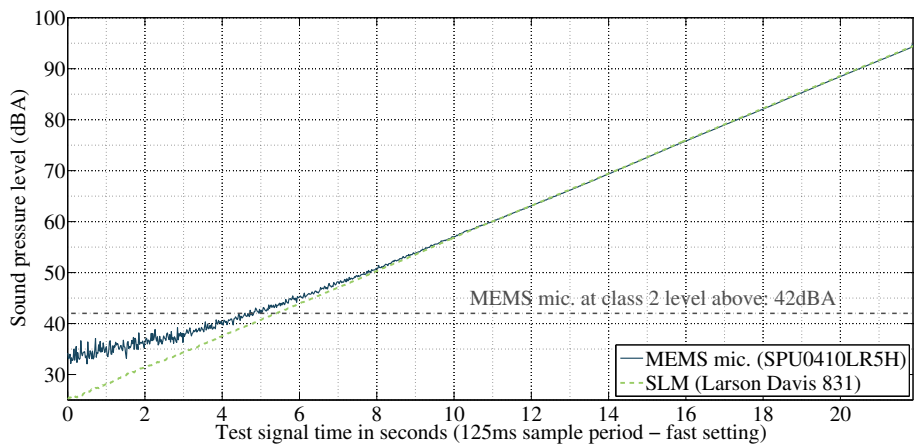}
\end{center}
\caption{\label{fig:level_resp}Linear level response of DUT vs. SLM to 1kHz sine wave upto 94dBA showing adjusted type 2 tolerance point.}
\end{figure}

For illustration, the vertical dashed line in Figure~\ref{fig:level_resp} shows the point at which the DUT meets the adjusted type 2 tolerance level ($\pm$0.6dB) for a 1kHz sinusoidal signal. The DUT can effectively operate within type 2 level linearity tolerances above \textbf{40dBA} on average for frequencies ranging from 31.5Hz - 8kHz. This lower limit can be reduced through the use of a lower noise microphone and pre-amp combination, as discussed in Section~\ref{sec:hardware_dev}, however this lower limit would rarely be observed in the urban sound environment. The DUT was also subjected to a linearly increasing pink and white noise signal, where the type 2 lower limit was observed at \textbf{37.2dBA} and \textbf{36.6dBA} respectively, highlighting the device's broadband linear response to varying urban SPLs.

\subsection{Toneburst response \label{sec:tb-resp}}
To test the DUT's response to transient SPLs, it was subjected to 4kHz sinusoidal tonebursts, varying in duration from 1000ms down to 0.25ms. IEC 61672-1, 5.9 defines tolerance limits in terms of dBA readings relative to the steady state 4kHz reading for type 2 devices. As these are relative measurements and do not rely on the use of the SLM as a reference, the type 2 tolerance limits as documented in IEC 61672-1 will be used.

\begin{table}
\centering
\small
\renewcommand{\arraystretch}{0.75}
\begin{tabular}{cccc}
\hline
Duration (ms) & IEC61672 $\Delta$ & DUT $\Delta$ & Tol.\\
\hline
1000 & 0.0 & 0.0* & $\pm$ 1.0 \\

500 & -0.1 & 0.0* & $\pm$ 1.0 \\

200 & -1.0 & 0.0* & $\pm$ 1.0 \\

100 & -2.6 & -2.0* &  $\pm$ 1.0 \\

50 & -4.8 & -4.0* & +1.0;-1.5 \\

20 & -8.3 & -7.9* &  +1.0;-2.0\\

10 & -11.1 & -10.9* &  +1.0;-2.0 \\

5 & -14.1 & -14.0* &  +1.0;-2.5 \\

2 & -18.0 & -18.4* &  +1.0;-2.5 \\

1 & -21.0 & -21.9* &  +1.0;-3.0 \\

0.5 & -24.0 & -25.7* &  +1.0;-4.0 \\

0.25 & -27.0 & -30.8* &  +1.5;-5.0 \\
\hline

\end{tabular}
\caption{Toneburst tests at 4kHz, varying duration (* indicates IEC61672-1, 5.9 type 2 criteria met)\label{tab:clause_18}}
\end{table}
As shown in Table~\ref{tab:clause_18}, the DUT met all IEC 61672-1, 5.9 criteria for 4kHz toneburst response. 

\subsection{Urban audio reproduction \label{sec:uar}}
\begin{figure}[h]
\begin{center}
\includegraphics[width=1.0\textwidth]{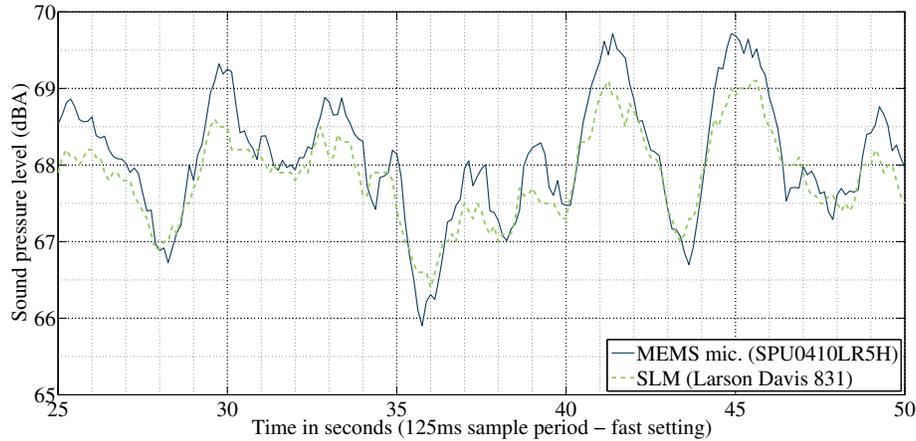}
\end{center}
\caption{\label{fig:urban_resp}Sample of DUT vs. SLM time history SPL values of urban audio recording reproduced under anechoic conditions}
\end{figure}

To further assess the DUT's ability to capture meaningful SPL data, a 15min urban audio recording was replayed a total of 4 times under anechoic conditions with the SLM and DUT microphone mounted directly adjacent to each other on-axis to the speaker. One of the more eventful samples of this time history data collected from the DUT and SLM is shown in Figure~\ref{fig:urban_resp}. It contains numerous impulsive events such as door closures and banging sounds.

As can be seen in Figure~\ref{fig:urban_resp}, the DUT closely follows the measurements made by the type 1 SLM. Correlation analysis was carried out on the resultant averaged SPL time histories from the SLM and DUT. The correlation coefficient ($R^2$) was calculated between the entire dBA (fast time weighting) time history for each device. The total $R^2$ value for this 15min urban signal was \textbf{0.9723} ($p\leq0.0001$). The mean difference between the SLM and DUT time history values was \textbf{0.4dB}, with a standard deviation of 0.1dBA, minimum vales of 0.1dBA and maximum values of 1.8dBA.

It seems that the MEMS microphone system slightly over-estimates the dBA values on the rise portion of transient sound events and slightly under-estimates on the  falling edge of these. This may be due to the fact that the DUT samples more frequently than the SLM resulting in this "over/under shooting" when measuring transient events.

\section{Future work \label{sec:conclusions}}
\subsection{Further measurements}
The full IEC 61672-1 standard includes specifications for parameters including: device directivity, high level thresholds and environmental variations, which require the full housing of the device to be incorporated. The final prototype will be tested against the extended set of requirements, including a long term exterior comparison against a type 1 SLM. Other factors such as the location of the sensor will be investigated as the majority of potential deployment locations are in close proximity to building facades. The resilience of these MEMS microphones to the varying environmental conditions of NYC is a critical aspect of this research. Further environmental testing is needed to quantify the effects of temperature and humidity on the devices response. Measurements will be carried out using equipment supplied by the Brookhaven National Laboratories, Biological, Environmental \& Climate Sciences Department~\footnote{\url{http://www0.bnl.gov/ebnn/becs/}} to test sensor functionality at extreme temperatures and humidities ranging from -20$^{\circ}$C to +50$^{\circ}$C and 25\% RH to 100\% RH. This will allow for the determination of sensitivity and frequency response variation under these varying conditions in a controlled environment.

\subsection{Hardware development\label{sec:hardware_dev}}
The high level RFI conditions in the vicinity of the sensor node and noisy low-cost power supply rely on a microphone solution with adequate RF shielding and a high power supply rejection ratio (PSRR). A digital MEMS microphone solution ensures that both of these external influences are no longer an issue when it comes to the gathering of high quality acoustic urban data. Noise observed on the output from the analog MEMS board is caused in part by parasitic noise from the power supply unit (PSU). This can cause measurement inaccuracies at particular frequencies where the noise is prevalent. The next iteration of the sensor's microphone solution will be an entirely digital design, utilizing a digital MEMS microphone (includes a built in ADC) and a USB audio CODEC enabling it to connect directly to the sensors computing device. The vastly improved power supply rejection ratio (PSRR) values and reduced EM/RF interference of the digital MEMS microphones over their analog counterparts should result in a much lower noise floor and an increase in dynamic range. The elimination of this noise will also result in an improved ability to capture clean audio signals for further in-situ processing and analysis.

The microphones non standard form factor is also worth revising. If a MEMS microphone could be built onto a circular 1/2inch PCB, the device could be calibrated using a standard 1/2inch acoustic calibrator making the calibration process much easier and potentially more accurate across multiple sensor nodes.

Battery powered sensor node solutions will also be investigated including power mode cycling and adaptation for periods of low acoustic activity.

\subsection{Automatic sound source identification \label{sec:ssid}}
The sensor presented in this article allows for the accurate, continuous monitoring of sound levels across a city. Whilst the gathering of accurate SPL data in-situ is crucial to the monitoring of noise in smart cities, identifying the source of these noise events is of great importance. The sensor's powerful processing unit means there is the capability of performing additional analysis of the audio signal. In tandem with the sensor development, considerable efforts have been employed on machine listening algorithms for the automatic identification of urban sound sources \cite{Salamon:UnsupervisedUrban:ICASSP:15,Salamon:ScattteringUrban:EUSIPCO:15}. One of the key advantages of running these classification models directly on the sensing device is that there is no need to transmit audio data to a centralized server for further analysis, in this way abating possible security and privacy concerns related to the recording of audio data. However, porting these models to the device presents a challenge due to the models' high computational complexity. To this end, future work will involve research into \emph{model compression} \cite{Ba:ReallyDeep:NIPS:14}, which can be used to obtain the performance of deep learning architectures using shallow ones which require less computational resources.

\section{Conclusion}
An advanced and accurate, low-cost sensor network has been presented. Based on this preliminary testing phase including the frequency compensation procedures, our analog MEMS microphone solution can produce accurate SPL data of high quality. Its adherence to the type 2 specifications for the tests undertaken is promising for its future use in a low-cost environmental acoustic sensor. The main limiting factor of its noise floor means it cannot effectively operate in ambient conditions of $<$30dBA, or at type 2 accuracies at levels $<$40dBA, however, this level would rarely be observed in the urban sound environment of NYC. The capabilities of this solution allow it to generate real-time acoustic data at or above the type 2 level. An accurate source of data from a reliable and low-cost sensor network is the cornerstone of any effective cyber-physical system for noise monitoring. With city agencies such as the NYC DEP relying on a minimum of type 2 level acoustic data, this solution can provide a reliable stream of data to inform and effectively prioritize existing noise enforcement procedures.

\section{Acknowledgments}
The authors would like to thank the New York University, Center for Urban Science and Progress (CUSP) for their seed funding of the current noise sensing initiative, Brookhaven National Labs for their equipment loan and support, The Cooper Union for the use of their acoustic laboratory, the Department of Environmental Protection (DEP) for their input on the NYC noise code and Deckard Audio for their input on the digital microphone design.


\bibliography{applied-acoustics}

\end{document}